# Accelerating Non-volatile/Hybrid Processor Cache Design Space Exploration for Application Specific Embedded Systems

Mohammad Shihabul Haque, Ang Li, Akash Kumar, Qingsong Wei*
National University of Singapore and *Data Storage Institute (DSI) Singapore
{matmsh, angli, akash} @nus.edu.sg, *WEI_Qingsong@dsi.astar.edu.sg

**Abstract**— In this article, we propose a technique to accelerate non-volatile/ hybrid of volatile and non-volatile processor cache design space exploration for application specific embedded systems. Utilizing a novel cache behavior modeling equation and a new accurate cache miss prediction mechanism, our proposed technique can accelerate NVM/hybrid FIFO processor cache design space exploration for SPEC CPU 2000 applications up to 249 times compared to the conventional approach.

## 1. INTRODUCTION

Presence of Non-Volatile Memory (NVM) cell in processor cache is no longer a science fiction. In the June of 2014, Toshiba Corporation revealed their non-volatile Perpendicular STT-MRAM cell based L2 processor cache that overwhelmed all the advantages of conventional volatile SRAM caches [20]. Toshiba Corporation also offer budget friendly STT-MRAM+SRAM hybrid cache [19]. Several commercially implementable designs came out recently to use non-volatile Phase Change Random Access Memory (PCRAM) cells with SRAM cells for low power hybrid processor caches (e.g. [17]). Moreover, extensive researches are going on to utilize the other NVM cell technologies (such as Resistive Random Access Memory, NAND Flash, etc.) in different levels of the processor cache hierarchy. The features that made NVM cells so attractive for processor caches; especially in energy-performance-area critical embedded systems, are (i) Higher storage density [13], (ii) Lower area occupancy [22] and (iii) Significantly lower energy consumption [14] over SRAM cells.

Besides being energy-performance-area critical, embedded systems are usually application specific. For application specific systems, in SRAM cache design space exploration [1], trace-driven single-pass cache simulators are widely used to quickly find the total number of cache misses during execution of the application on different cache configurations [24, 9]. We call this phase as the cache performance evaluation phase. Once the cache performance (i.e. number of cache misses) is known, analytical models (such as the one in [12]) can be used to calculate the amount of energy and area consumption by each cache configuration simulated. The cache configuration that best suits the performance-energy-area occupancy criteria is chosen for the final system design. Available single-pass cache simulators are ill-suited to fulfill the requirements of the cache performance evaluation phase in NVM/hybrid cache design space exploration. Two such requirements are discussed below:

- Almost every type of NVM cell is prone to wear out when written heavily (i.e. loaded with data/updated frequently) [15, 29]. Therefore, in a heavily written cache that deploys NVM cells/cache lines (from here we use the words "cell" and "cache line" interchangeably) with limited write endurance, cache configuration as well as performance and energy consumption may change over time due to cache line wear out. Change in cache performance and energy consumption can be fatal in systems such as application specific real-time embedded systems. Therefore, the system designer must deploy a safety mechanism to prevent the system from being used when its cache configuration reaches to an unsuitable state. To assist the system designer in designing a safety mechanism or to identify the unsuitable states in a cache during design space exploration, the cache performance evaluation phase needs to evaluate the performance in every initially deployable cache configuration as well as in every other configuration it generates due to line wear out.

- Besides limited write endurance, any type of NVM cell is far more expensive than SRAM cell. Moreover, writing time is significantly slower in some types of NVM cells compared to SRAM cells (e.g. PCRAM write latency is 5ns where SRAM write latency is only 1ns [17]). Due to one or few of these reasons, instead of deploying NVM cell in every cache line, hybrid cache that deploys different types of cells (such as PCRAM, STT-MRAM, SRAM, etc.) is used. To meet budget, cache performance and/or cache lifespan constraints, a hybrid cache should deploy different types of cells in such a combination that (i) price of the cache remains reasonable, (ii) application executes at a reasonable speed, (iii) NVM cells/lines are not used in the heavily written lines, and/or (iv) cache performance and energy efficiency does not degrade significantly when couple of NVM lines wear out. To assist in finding the most suitable combination of cells for a hybrid cache configuration during design space exploration, the cache performance evaluation phase needs to report the number of writes per cache line and per cache set.

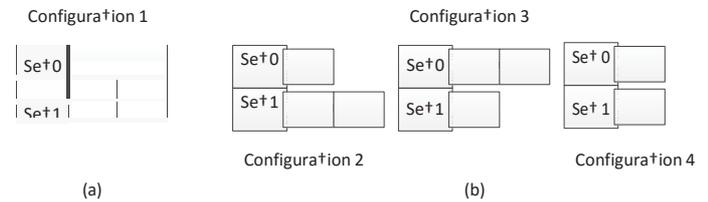

Figure 1: An NVM Cache's Configurations Due to Line Wear Out

Let us explain with an example how these new requirements make the available single-pass cache simulators infeasible for NVM/hybrid processor cache design space exploration. Figure 1 (b) shows all the three additional configurations that may generate from the NVM/hybrid FIFO cache configuration shown in Figure 1 (a) when the cache becomes unusable after the last active line in any cache set wears out. That means, if the set size was 8 and associativity was 64 in the cache configuration of Figure 1 (a), it was necessary to simulate 281,474,976,710,655 configurations in the existing single-pass cache simulators. Simulating so many configurations will take few years for sure by the exiting single-pass cache simulators. Moreover, to record the number of misses per line in each of these cache configurations, enormous amount of storage is required.

In this article, we propose a trace-driven resource generous technique "Breakneck Cache Performance Evaluation Method" ("BCPEM") to accelerate the cache performance evaluation phase in NVM/hybrid processor cache design space exploration for application specific embedded systems. "BCPEM" is exclusively designed for First-In-First-Out (FIFO) replacement policy as FIFO is very popular in embedded processor caches [9]. "BCPEM" can accelerate efficiently as long as (i) line wear out in one cache set does not influence the number of misses in other cache sets in an NVM/hybrid cache and (ii) line size is same in all the cache configurations to analyze. To accelerate, "BCPEM" utilizes the following approaches:

- Instead of simulating each cache configuration, "BCPEM" evaluates performance of each cache set separately. For example, if the

---
[1] The process of finding the most suitable processor cache configuration. Combination of the following cache parameters is defined as cache configuration: (i) set size: number of cache sets, (ii) Associativity: number of storage cells/cache lines per set, (iii) Line size: amount of data storable in each storage cell, and (iv) Replacement Policy: policy to select a storage cell to load new data

*Therefore, in equation 1,*

$$n = (V_1 \times (n^* + n_0)) - n_0$$
$$\text{or, } n_0 = (V_1 \times (n^* + n_0)) - n$$
$$\text{or, } n_0 = (V_1 \times n^*) + (V_1 \times n_0) - n$$
$$\text{or, } n_0 \times (1 - V_1) = (V_1 \times n^*) - n$$
$$\text{or, } n_0 = \frac{(V_1 \times n^*) - n}{1 - V_1} \quad (2)$$

*By replacing $V_1$ with $(1/2) \times (H + \sqrt{H^2 - 4})$ in equation 2,*

$$n_0 = \frac{((1/2) \times (H + \sqrt{H^2 - 4}) \times n^*) - n}{1 - ((1/2) \times (H + \sqrt{H^2 - 4}))} \quad (3)$$

Equation 3 suggests, *as $H$ and $n$ are variables dependent on $M$, and $M_0$ is fixed for the given application, we can never find a value for $n_0$ that does not change with the change in $M$. Hence, $n_0$ should be replaced with a variable $R_0$ dependent on associativity ($M$).*

Involvement of $M_0$ in $H$ raised the second concern in our analysis. No cache line is considered to be blocked in the available trace-driven cache simulators [6, 12, 24]. Therefore, even without $M_0$, the modeling equation should be able to estimate the total number of cache misses ($n$) for a given associativity ($M$). With these changes, the equation to model the effect of $M$ on $n$ should look like the following:

$$n = (1/2) \times (H + \sqrt{H^2 - 4}) \times (n^* + R_0) - R_0 \quad (4)$$

where $H = 1 + \frac{(M)}{(M^*)}$ and $M < M^*$. We named equation 4 as the "Breakneck Modeling Equation" ("BME"). like "PFE", "BME" is also not efficient in modeling for a cache configuration on which an application's $n$ is close to $n^*$ (i.e. difference between $n$ and $n^*$ is less than 20%) and/or 80% or more memory accesses by the processor generate cache miss.

## 4. BREAKNECK PREDICTION METHOD

When associativity decreases in a fully-associative cache/set, usually number of misses ($n$) increases for a particular application. If the number of misses ($n$) for associativities $M_x, M_{x-1} = M_x - 1, M_{x-2} = M_{x-1} - 1, ... M_1$ are $n_1, \frac{n_1}{2^{x-2}}, \frac{n_1}{2^{x-3}}, ... n_1$, the $n$s for $M_x$ to $M_1$ are certainly not collinear. Similarly, if the number of misses ($n$) for associativities $M_x$ to $M_1$ are $\frac{n_1}{1.5^{x-1}}, \frac{n_1}{1.5^{x-2}}, ... n_1$, the $n$s for $M_x$ to $M_1$ are not collinear; however, the curve generated by the $n$s is more linear compared to the previous case. In fact, it is very easy to verify using Microsoft excel or similar software that, to be considered collinear, $n$s for all the associativities in between and including $M_x$ and $M_1$ must not differ by more than 9% compared to the $n$ of $M_x$. **When collinear, by knowing $n$s for associativities $M_x$ and $M_1$ only, it is possible to calculate/predict $n$s for all the associativities in between $M_x$ and $M_1$ using linear interpolation. This method can save a huge amount of time in the cache performance evaluation phase.**

To find the range of associativities that have collinear $n$s (such as $M_x$ to $M_1$), we can use "BME". For all the associativities in between and including $M_x$ and $M_1$, if $H$s are very close (i.e. does not differ significantly compared to the $H$ of $M_x$) and can be considered equal/fixed/constant, then equation 4 can be written as

$$n = C_1 \times (n^* + R_0) - R_0$$

where constant $C_1 = (1/2) \times (H + \sqrt{H^2 - 4})$

due to line wear out, in the target cache set for the given application. After that, a single-pass cache simulator have to be used to simulate those pairs of $M$s within which $H$s of any $M$ does not change by 20% or more compared to the $H$ of the largest $M$ in the pair [2]. If for any simulated associativity pair $\{M_i, M_j\}$, there exists no simulated associativity $M_k$ such as $M_i < M_k < M_j$, and $n$s for $M_i$ and $M_j$ varies by 10% or more, a new associativity $M_k = \frac{(M_i + M_j)}{2}$ has to be simulated. This process is repeated unless between simulated associativities $M_i$ and $M_j$, there exists no simulated associativity $M_k$, and $n$ for $M_i$ is not larger than 9% or more compared to the $n$ for $M_j$. For example, assume that the single-pass cache simulator simulated associativities ($M$) 1, 16 and 44 for set 0 in a cache with four sets and cache line size 16 bytes. It means, between associativity 1 and 16 (as well as between associativities 16 and 44) $H$s do not change by 20% or more. If $n$ for associativity 1 is 10% larger than associativity 16, simulator needs to simulate associativity 8. If $n$ for associativity 1 is 10% larger than associativity 8, simulator needs to simulate associativity 4 and so on. When between any simulated associativities $M_i$ and $M_j$, there exists no simulated associativity $M_k$, and $n$s for $M_i$ and $M_j$ varies by less than 10%, linear interpolation can be used to calculate/predict the number of cache misses in all the associativities in between $M_i$ and $M_j$. As a result, simulation time can be reduced significantly. We name this prediction method as "Breakneck Prediction Method" ("BPM").

A point to note, instead of linear interpolation in "BPM", "BME" can be used to predict the $n$s for all the associativities in between $M_1$ and $M_x$ when their $n$s are collinear. However, use of "BME" to predict $n$ is slightly time consuming compared to linear interpolation. This is because, to predict the $n$s using "BME" for all the associativities in between $M_1$ and $M_x$, $R_0$s of $M_1$ and $M_x$ have to be calculated using "BME" first. After that, using $R_0$s of $M_1$ and $M_x$ in regression analysis or linear interpolation (as $R_0$s for associativities $M_x$ to $M_1$ should be collinear according to Equation 5), $R_0$s for all the associativities in between $M_1$ and $M_x$ have to be approximated/predicted. When $n^*$, $H$ and $R_0$ for an associativity ($M$) are known for a given application, "BME" can be used to predict/estimate the number of cache misses for that $M$ to avoid simulation.

## 5. BREAKNECK CACHE PERFORMANCE EVALUATION METHOD

Breakneck Cache Performance Evaluation Method ("BCPEM") is an application specific, trace-driven technique which is aimed to speed up the cache performance evaluation phase in NVM/hybrid set-associative FIFO processor cache design space exploration, when line wear out in one cache set does not influence the number of misses in other cache sets. "BCPEM" utilizes the fact, when line wear out in a cache set does not influence the number of cache misses ($n$) in other cache sets, every cache set can be considered as an independent fully-associative cache. That means, in Figure 1 (b), set 0 in both Configuration 2 and Configuration 4 will generate the same number of cache misses for a given application. Therefore, by collecting $n$ in set 0 associativity 1, set 1 associativity 1 and set 1 associativity 2 separately, and by combining the results, $n$ of Configuration 2 and Configuration 4 can be estimated accurately. It is an wastage of time to collect $n$ of set 0 in both of Configuration 2 and Configuration 4. Due to this wastage of time, existing single-pass cache simulators are infeasible to deploy when the NVM/hybrid cache has large number of sets and each set can generate large number of different associativities due to line wear out.

"BCPEM" cache performance evaluation flow is presented in Figure 2.

"BCPEM" collects $n$ of each cache set in a cache for different associativities that may generate due to line wear out. For this purpose, "BCPEM"

$$\text{or, } n = (C_1 \times n^*) + (C_1 \times R_0) - R_0,$$
$$\text{or, } n = (C_1 - 1) \times R_0 - C_2 \text{ where constant } C_2 = (C_1 \times n^*)$$
$$\text{or, } n = C_3 \times R_0 - C_2 \text{ where constant } C_3 = (C_1 - 1)$$
(5)

Equation 5 is a straight line equation and, therefore, ns for $M_x$ to $M_1$ are collinear. However, the question remains, when can we consider the Hs for associativities $M_x$ to $M_1$ as constant? Even though we do not know the answer, the following two information are enough to design an efficient cache miss prediction method: (i) Hs for $M_1$ to $M_x$ must not differ significantly, and (ii) ns for $M_1$ to $M_x$ must not differ by more than 9% compared to the n of $M_x$.

Now, let us explain how the prediction method should work. Initially, it is necessary to calculate H values for all the associativities (M) possible, collects $n^*$ and $M^*$ for each cache set first. After that, for each cache set, "BCPEM" collects n for the associativities possible to generate due to line wear out, by simulating only few of those associativities. Let us explain these steps in details:

**Step1:** $n^*$ *calculation* - Total number of cold misses ($n^*$) for a given application in a cache set is the total number of unique data blocks loaded in that cache set. A cache memory loads data blocks containing multi-set size is 8 and associativity is 64 in a cache initially, "BCPEM" evaluates performance of each of the 8 cache sets separately for associativities 64, 63, 62,...1 when line wear out can reduce associativity to 1. Therefore, only 512 *cache set* configurations are necessary to evaluate rather than 281,474,976,710,655 *cache configurations*. This approach reduces storage requirement significantly.

---

[2] Definition of H in Section 3 suggests, when maximum associativity is 64 and optimal associativity is up to 10 million, 20% difference between the Hs of $M_1$ and $M_x$ can help to skip simulation of up to ten consecutive associativities. It is not a good idea to skip simulation of more than 10 consecutive associativities for which we want to predict ns using linear interpolation

- Per simulated cache set, by simulating few of the associativities,

"BCPEM" predicts the number of cache misses for all the associativities possible due to line wear out. "BCPEM" utilizes "Breakneck Modeling Equation" ("BME") and "Breakneck Prediction Mechanism" ("BPM") to predict the number of misses in the associativities which are not simulated. Once the number of cache misses is known for a given application for each of the associativities possible (e.g 64 to 1) in each cache set (e.g. each of the 8 sets), "BCPEM" can calculate the number of cache misses in all the cache configurations possible in the NVM/hybrid cache (e.g. total of 281,474,976,710,656) in a split of a second.

This article discusses "Breakneck Cache Performance Evaluation Method" ("BCPEM") as follows: Section 2 discusses some related works; Section 3, 4 and 5 discuss "BME", "BPM", and "BCPEM" respectively, Section 6 shows the efficiency of "BCPEM" with empirical evidences and Section 7 concludes the paper.

Throughout the article, we made the following assumptions:
- One faulty bit makes a cache line completely unusable.
- When all the lines wear out in a set, the set as well as the entire cache dies/wears out.
- Line size is same in all the cache configurations.

## 2. RELATED WORK

Among the available cache simulation techniques (such as system simulation [28], instruction set simulation [16], etc.), application's memory access trace-driven single-pass cache simulation is known to be the fastest and the most resource generous. In a single-pass cache simulator, a trace file that indicates when and which data blocks were accessed by the processor during execution of an application is used as the input. By reading one data block access at a time from the trace file, the single-pass simulator checks whether the requested data block is available in the simulated cache configurations. Cache configurations are represented by an array or a list in single-pass simulation. Therefore, without spending a large amount of time in simulating the exact hardware behavior (unlike system simulation [28] and instruction set simulation [16]), single-pass cache simulators can quickly and accurately estimate the number of cache misses for a particular application on a group of cache configurations.

To mimic the hardware behavior minimally and to reduce the need for extensive computing resources, additional mechanisms, such as special data structures [8], trace compression [18, 27], running the simulation on parallel hardware [10, 25], inclusion properties [24], etc. are applied in single-pass simulation (Point to note, not all of these mechanisms can be used for every cache replacement policy). The state-of-art, single-pass FIFO cache simulator is "CIPARSim" [9]. Like the other existing cache simulators, "CIPARSim" is also not fast enough to fulfill the requirements of NVM/hybrid FIFO processor cache design space exploration for application specific systems.

Modeling the effect of caching capacity on cache misses is mathematically challenging. However, having such an analytical model can help to estimate the number of cache misses during an application without performing any time consuming simulation. Therefore, such an analytical model can be the perfect choice to replace single-pass cache simulators in the cache performance evaluation phase during NVM/hybrid FIFO processor cache design space exploration. Analytical modeling possibilities have been studied widely but without much success. Earlier works in this attempt found the necessity to adapt simple memory access models [1, 21] and/or focus on specific replacement policies [7, 4]. However, they neither take into account the interaction among processes and with the kernel [11], nor the changes in the memory access pattern [3]. Considering the previous limitations, Tay et al. [23] proposed the following analytical model:

$$n = (1/2) \times (H + \sqrt{H^2 - 4}) \times (n^* + n_0) - n_0 \quad (1)$$

where $H = 1 + \frac{(M^* - M_0)}{(M - M_0)}$ and $M < M^*$. In this equation, $n$ is the caching capacity $M$ (for which the number of misses has to be estimated) changes. $M^*$ is the optimal caching capacity of a fully-associative cache that does not require any reloading of any data block for a given application. $M_0$ represents the amount of cache space that cannot be used due to some sort of locking mechanism or scenario-specific reasons. $n^*$ is the number of cold misses for an application. $n_0$ is a correction factor independent of $M$. $M^*$, $M_0$, $n^*$ and $n_0$ are fixed for a given application, operating system and hardware. **This modeling equation is not efficient for a cache configuration on which an application's $n$ is close to $n^*$ or cache hit rate is too low.**

Equation 1 is called "Page Fault Equation" (referred to as "PFE"). This equation can model the effect of caching capacity on the number of cache misses efficiently (for real workloads on Linux and Windows 2000 that include different replacement policies, compute, IO and memory intensive benchmarks, multiprogramming and user input. Performance of "PFE" has never been evaluated for SPEC CPU 2000 and 2006 benchmark applications) when the values of $M^*$, $M_0$, $n^*$ and $n_0$ are provided for an application. To obtain/approximate the values of $M^*$, $M_0$, $n^*$ and $n_0$, simulations are performed to know the number of misses for all the caching capacities (Ms) possible in a fully-associative cache and the results are used in regression analysis. The less number of caching capacities are simulated, the less the efficiency of "PFE". Therefore, "PFE" is not helpful enough to be utilized as a cache miss prediction method to avoid cache simulation in the cache performance evaluation phase in design space exploration. In addition, as there was no known fast method to calculate the actual values of $M^*$ and $n^*$, it was impossible to verify the definition of $M^*$ and $n^*$ used in "PFE" and accuracy of their values found via regression analysis. Moreover, "PFE" does not consider cache set size, and no proposal was made on how to adapt "PFE" on modern set-associative processor caches. Therefore, this equation could not help much either in SRAM or in NVM/hybrid cache design space exploration.

Several system-level analytical models are available to collect information for architectural studies such as the effect of associativity on the memory fetch time, effect of cache line size on cycle time, etc. CACTI [26] is one such famous system-level model for SRAM caches. For NVM, NVSim [5] is a popular system-level model based on CACTI. However, these system-level analytical models are not capable to model the effect of associativity or caching capacity on the number of cache misses for a given application.

## 3. BREAKNECK MODELING EQUATION

Realizing the potential of analytical modeling and inspired by the success of "PFE" [23], we became interested to analyze "PFE" in search for a modeling equation that can speed up the cache performance evaluation phase in NVM/hybrid FIFO processor cache design space exploration. Here, we discuss our findings that lead us to "Breakneck Modeling Equation" ('BME'). BME is the basis of "BCPEM".

To have a better understanding, we started with verifying the definition of each parameter and variable used in "PFE". As "PFE" does not take cache set size into account, it was intuitive that "PFE" should be applicable to each set separately in a set-associative cache. Therefore, $M$ **should be the associativity and $M^*$ should be the optimal associativity to avoid reloading of any data block for a given application.** We also designed our own algorithm to find the actual values of $M^*$ and $n^*$ quickly and accurately (discussed in Section 5). With the set-associative cache specific definition of $M$ and actual values of $M^*$ and $n^*$, we discovered that, even when $n$s for all the possible $M$s are known and used in regression analysis for "PFE", the approximate values do not quite match with the actual values of $n^*$ and $M^*$. Moreover, when we provide the actual values of $M^*$ and $n^*$ in regression analysis to approximate $n_0$ and $M_0$ only, "PFE" modeling efficiency degrades. It indicates a problem with the definition or use of $n_0$, $M_0$, $M^*$ and $n^*$ in "PFE".

To identify the root of the problem, we decided to analyze $n_0$ as the starting point. We noticed that $n_0$ changes significantly when the number of known data points (associativities for which cache misses are known via simulation) in regression analysis changes. It suggests that $n_0$ **should be replaced with a variable $R_0$ which is dependent on associativity** (i.e. when $M$ changes, $R_0$ value should change for a given application. In the original proposal of PFE, $n_0$ was fixed for a given application). Below, we provide a mathematical proof to support our claim.

*Proof Against Fixed $n_0$ in "PFE"*: In equation 1 of Section 2, $(1/2) \times$ total $(H + \sqrt{H^2}$

$- 4) = V_1$ *is a variable dependent on associativity* $M$ *and number of cache misses and it varies when the fully-associative cache's* $n$ *is fixed/constant for the given application.*

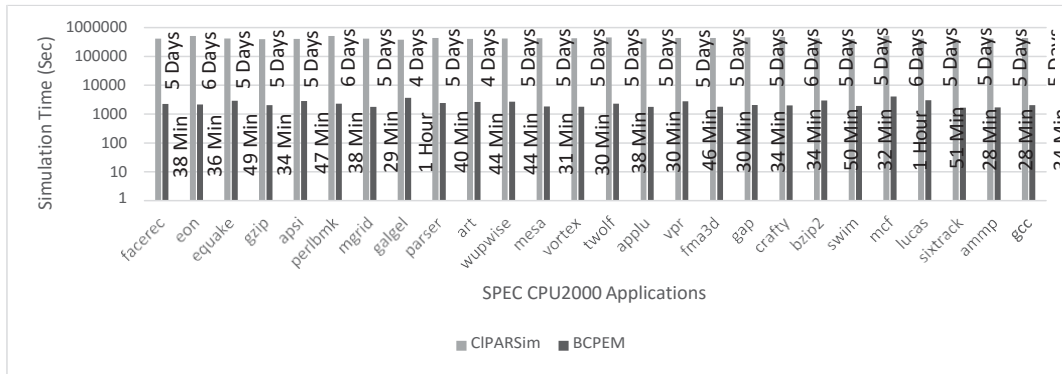

Figure 4: Entire Cache Performance Evaluation Time for SPEC CPU 2000 Applications

to 8 (for application bwaves) times faster than "CIPARSim". "BCPEM" could do the job so quickly because, cache misses were predicted for a large number of associativities using "BPM". Due to the use of "BPM", "BCPEM" will always be faster than "CIPARSim" no matter how small the application is.

The first two parts of our experiment made it evident that, as (i) the same cache simulator is used to simulate less and simpler configurations (i.e. cache set rather than an entire cache) and (ii) $n^*$ and $M^*$ estimation time is ignorable, time consumed by "BCPEM" can never exceed the time consumed by "CIPARSim". Therefore, in the third part, to figure out how much time can be reduced by "BCPEM" compared to "CIPARSim" for an entire benchmark suite, we chose SPEC CPU 2000 suite. To complete our experiment in reasonable time, we used (i) the cache configuration of the previous experiment but with reduce line size of 4bytes and (ii) trace of the first 100 Million instructions for each application. Figure 4 compares the total time consumed by "BCPEM" and "CIPRASim" to collect n of each SPEC CPU 2000 application on the 3,249 configurations possible in the target cache. For these applications too, "BCPEM" is 104 (for application "galgel", "BCPEM" took 1 hour and "CIPARSim" took 4 days) to 249 (for application "ammp", "BCPEM" took 28 minutes and "CIPAR-Sim" took 5 days) times faster than "CIPARSim". For SPEC CPU 2000 applications, accuracy of predicted misses in "BCPEM" were within +6% and -5%. Therefore, "BCPEM" is an efficient acceleration technique for NVM/hybrid FIFO processor cache design space exploration.

## 7. CONCLUSION

Trace-driven single-pass cache simulators are not as successful in accelerating non-volatile/hybrid of volatile and non-volatile processor cache design space exploration as they are in SRAM processor cache design space exploration for application specific embedded systems. In this article, we propose a technique to accelerate the phase which is conventionally accelerated by single-pass cache simulators in processor cache design space exploration for application specific embedded systems. Utilizing a novel cache behavior modeling equation and a new accurate cache miss prediction mechanism, our proposed technique can accelerate NVM/hybrid FIFO processor cache design space exploration for SPEC CPU 2000 applications up to 249 times compared to the conventional approach.

This work was supported by Agency for Science, Technology and Research (A*STAR), Singapore under Grant No. 112-172-0010.